\documentclass[prd,aps,floatfix,nofootinbib,preprint ,tightenlines,superscriptaddress]{revtex4}
\usepackage{dcolumn}
\usepackage{amsmath}
\usepackage{latexsym}
\usepackage{graphicx}
\usepackage{bm}

\def\svev#1{\left\langle #1\right\rangle}       

\def\Tr{{\rm Tr}\,}

\newcommand{\bee}{\begin{equation}}
\newcommand{\ee}{\end{equation}}
\newcommand{\beea}{\begin{eqnarray}}
\newcommand{\eea}{\end{eqnarray}}

\begin{document}

\title{
Simple chromatic properties of gradient flow
}
\author{Thomas DeGrand}
\email{thomas.degrand@colorado.edu}
\affiliation{Department of Physics, University of Colorado, Boulder, CO 80309, USA}

\begin{abstract}
It has become customary to use a smoothing algorithm called
 ``gradient flow'' to fix the lattice spacing in a simulation, through a parameter
called $t_0$. It is shown that in order to keep the length $t_0$ fixed with respect
to mesonic or gluonic observables as the number of colors $N_c$ is varied, the fiducial point
for the flow parameter must be scaled nearly linearly in $N_c$. In simulations with dynamical fermions, the
dependence of $t_0$ on the pseudoscalar meson mass flattens as the number
of colors rises, in a way which is consistent with large $N_c$ expectations.

\end{abstract}

\maketitle

\section{Introduction}

The predictions of lattice studies of systems like QCD are of dimensionless quantities,
such as the ratio of two masses. One often wants to present these results as dimensionful
numbers (such as masses in GeV). This is done by picking one observable as a fiducial, fixing its value
somehow to experiment, and
expressing all one's results in terms of it.
 In lattice QCD
simulations, many choices for a scale-setting parameter
 have been used \cite{Sommer:2014mea}: masses of various stable particles, decay constants, or quantities derived
from the heavy quark potential, such as the string tension
or of inflection points in the potential (Sommer parameters ~\cite{Sommer:1993ce}).
As long as one is studying some system in isolation, there is no deep reason 
(though there might be practical ones) to favor one 
choice for a parameter over another. Indeed, the most used quantities for scale setting are 
arbitrary choices with no direct connection to observation.

There are situations when one might want to compare different theories to each other.
The particular comparison, which is the subject of this note, is for
systems with different numbers of colors $N_c$. I am concerned with
the $N_c$ dependence of a new fiducial quantity, a squared distance conventionally labeled $t_0$,
 which is derived from the diffusive smoothing
of the gauge field  \cite{Luscher:2010iy,other}, through a process called ``gradient flow''
or ``Wilson flow.'' The use of $t_0$ to set the scale has become standard due to its high accuracy
and ease of use. There is a high probability that it will be adopted as a scale setting fiducial
for other confining and chirally broken systems. This short paper addresses two questions
related to the use of $t_0$ in such studies:

First, $t_0$ is a derived quantity; a certain gauge observable, to be defined below, is set to
some value which determines $t_0$. How should that value be set, so that the scale $t_0$
remains constant with respect to other scales set by gluonic or mesonic observables, as $N_c$ is varied?
A simple expectation will be given and tested.

Next, there is a prediction due to B\"ar and Golterman \cite{Bar:2013ora}, for the fermion mass dependence of $t_0$.
It comes from a chiral Lagrangian analysis and the small mass limit of their formula involves
the pseudoscalar mass $m_{PS}$, the pseudoscalar decay constant $f_{PS}$ and an undetermined constant $k_1$,
\bee
t_0(m_{PS}) = t_0(0)( 1 + k_1 \frac{m_{PS}^2}{f_{PS}^2} +\dots )
\ee
(the full formula is given in Eq.~\ref{eq:bg}, below). Essentially all large scale simulations which
measure $t_0$ observe the linear dependence of $t_0$ on $m_{PS}^2$, but with only one value of $N_c$ there 
is not much one can say about the $k_1/f_{PS}^2$ part of the expression.
Data at several values of $N_c$ reveal that $k_1/f_{PS}^2$ decreases as $N_c$ rises, in a way which is consistent with
large $N_c$ expectations.

In `t Hooft's \cite{'tHooft:1973jz} analysis of QCD in the limit of large number of colors,
 observables have a characteristic scaling with the number of colors $N_c$.
As in a lattice calculation, the most correct way to express these relations is to talk about
dimensionless ratios, though usually
this is expressed through statements like ``meson masses $m_M$ are independent of 
$N_c$, while decay constants  scale as $f_{PS}\sim\sqrt{N_c}$.'' I will use this language
in the text. Large $N_c$ expectations, which are well satisfied by lattice data (compare results from
pure gauge simulations, summarized in Ref.~\cite{Lucini:2012gg} as well as ones involving fermions from
Refs.~\cite{DeGrand:2012hd,Bali:2013kia,DeGrand:2016pur}), are that
when simulations are performed at the same values of the bare 't Hooft coupling $\lambda=g^2 N_c$,
mesonic observables and ones derived from the static potential are approximately 
independent of $N_c$, while other observables scale appropriately.

 ``Gradient flow''  or ``Wilson flow''   is a smoothing
method for gauge fields achieved by diffusion in a fictitious (fifth dimensional) time $t$.
 In  continuum language, a smooth gauge field $B_{t,\mu}$ is defined in terms of the original 
gauge field $A_\mu$ through an iterative process
\beea
\partial_t B_{t,\mu} &=& D_{t,\mu}B_{t,\mu\nu}  \nonumber \\
B_{t,\mu\nu} &=& \partial_\mu B_{t,\nu} - \partial_\nu B_{t,\mu} + [B_{t,\mu},B_{t,\nu}] , \nonumber \\
\label{eq:flow1}
\eea
where the smoothed field begins as the original one,
\bee
B_{0,\mu}(x)=A_\mu(x).
\ee
L\"uscher  \cite{Luscher:2010iy} proposed measuring a distance from flow, using the 
field strength tensor built using the $B's$, $G_{t,\mu\nu}$, via the observable
\bee
\svev{E(t)} = \frac{1}{4}\svev{G_{t,\mu\nu}G_{t,\mu\nu}}.
\label{eq:def}
\ee
The definition of a squared length $t_0$ comes from fixing the value of the observable
to some value $C(N_c)$
\bee
t_0^2 \svev{E(t_0)} = C(N_c)
\label{eq:flow}
\ee
and treating $t_0$ as the dependent variable.

Empirically, it is known that at very small $t$, $t^2 \svev{E(t)}$ rises quickly from zero, and then flattens into a 
linear function of $t$. The value of $C(N_c)$ which fixes $t_0$ is chosen to be some value in the linear region.

How does $C(N_c)$ vary with the number of colors, compared to other observables which are expected
to be independent of $N_c$? L\"uscher reports that,
perturbatively,
\bee
t^2 \svev{E} = \frac{3}{32\pi}(N_c^2-1) \alpha(q)[1 + k_1 \alpha + ...]
\label{eq:pertflow}
\ee
where $\alpha(q)$ is the strong coupling constant at momentum
scale $q \propto 1/\sqrt{t}$.
Using the one-loop formula for the coupling constant,
\bee
\frac{1}{\alpha(q)} = N_c \frac{B(N_c,N_f)}{2\pi} \log \frac{q}{\Lambda},
\label{eq:oneloop}
\ee
where $ B(N_c,N_f)= 11/3 - (2/3) N_f/N_c$, we invert Eq.~\ref{eq:pertflow} to find
\bee
\log \frac{q}{\Lambda} =  \frac{3}{16}\frac{1}{B(N_c,N_f)C(N_c)} [N_c + O(1)+O(\frac{1}{N_c}) + \dots].
\label{eq:scale}
\ee
The scale $q$  is an inverse distance. This expression  says that, in order to match distances
across $N_c$, in units of $\Lambda$, it must be that $C(N_c) = A_1 N_c + A_0 + ...$.
This formula is what I wish to test.

 Our scale setting observable is
$r_1$, the shorter version of the Sommer parameter \cite{Bernard:2000gd}. 
For ordinary QCD, $r_1=0.31$ fm \cite{Bazavov:2009bb}.
Its value for the data sets which will be displayed has been previously  published in
Refs.~\cite{DeGrand:2012hd,DeGrand:2016pur}.

 \section{Simulation details}

The data sets are the ones presented in Refs.~\cite{DeGrand:2016pur}  and \cite{DeGrand:2012hd}
plus some additional ones
to be described below. The simulations used the Wilson gauge action and clover fermions with
normalized hypercubic links~\cite{Hasenfratz:2001hp,Hasenfratz:2007rf}.
The dynamical fermion simulations had $N_f=2$ flavors of degenerate mass fermions.
All lattice volumes are $16^3\times 32$. The data sets were approximately matched
in lattice spacing, so not much can be said about the size of
discretization artifacts. (Note, however, that large $N_c$ comparisons do not necessarily have to be
made in 
the continuum limit.) The spectroscopic data sets were based on about 100 lattices per bare parameter
value.
(The precise numbers were given in Refs.~\cite{DeGrand:2016pur,DeGrand:2012hd}.) Table \ref{tab:t0}
records the number of lattices on which flow variables were measured.
The lattices from dynamical fermion data sets were typically separated by 10 molecular dynamics time
steps;
the quenched lattices were separated by 100 Monte Carlo updates using a mixture of over-relaxation and
heat bath.

The extraction of $t_0$ from lattice data is standard. The gradient flow differential equation is
 integrated numerically using
the Runge-Kutta algorithm generalized to $SU(N_c)$ matrices, as originally
proposed by L\"uscher \cite{Luscher:2010iy}.
 The routine discretizes the flow time with a step size $\epsilon$. Calculations used the usual
``clover'' definition of $E(t)$ \cite{Luscher:2010iy}.

Three aspects of the data need to be described, all of which could influence
the results. The first is the choice of integration step size $\epsilon$. To check this, I took one
data set (one $\kappa$ or bare quark mass value) 
per $SU(N_c)$ and generated an additional data set at a larger step size.
Specifically, the data in the  tables uses $\epsilon=0.03$ for $N_c=2-4$ and 0.05 for $N_c=5-7$. 
I augmented this with an $\epsilon=0.05$ data set for $N_c=2-4$ and $\epsilon=0.07$ at $N_c=5-7$.
Identical analysis on the two data sets revealed no differences between the results with the two values
of $\epsilon$ (or more precisely, the differences were about an order of magnitude smaller that the
quoted
 uncertainties). 

Next, the dynamical fermion data sets are presumably correlated in molecular dynamics simulation time.
I attempted to estimate the autocorrelation time through the autocorrelation function
 (for a generic observable $A$) defined as
\bee
\rho_A(\tau) = \frac{\Gamma_A(\tau)}{\Gamma_A(0)}
\ee
where
\bee
\Gamma_A(\tau)= \sum_{i=1}^N \svev{(A(\tau-\bar A)(A(0)-\bar A)}.
\ee
The integrated autocorrelation time (up to a window size $W$) is
\bee
\tau_{int}(W) = \frac{1}{2} + \sum_{\tau=1}^W \rho(\tau).
\ee
An issue with these observables is that unless the total length in time of the data set is much larger
than the auto-correlation time, it is difficult to estimate an error for them. That is a problem with
most
of the data sets used; there are typically $O(100)$ measurements. However, it happens that I have
additional data for several of the  $SU(3)$ and $SU(4)$  sets with about 5000 
equilibrated trajectories and 500 saved lattices.
I analyzed these sets sets by breaking them into five parts, computing $\tau_A$ on each part,
and taking an error from the part-to-part fluctuations.

\begin{figure}
\begin{center}
\includegraphics[width=0.8\textwidth,clip]{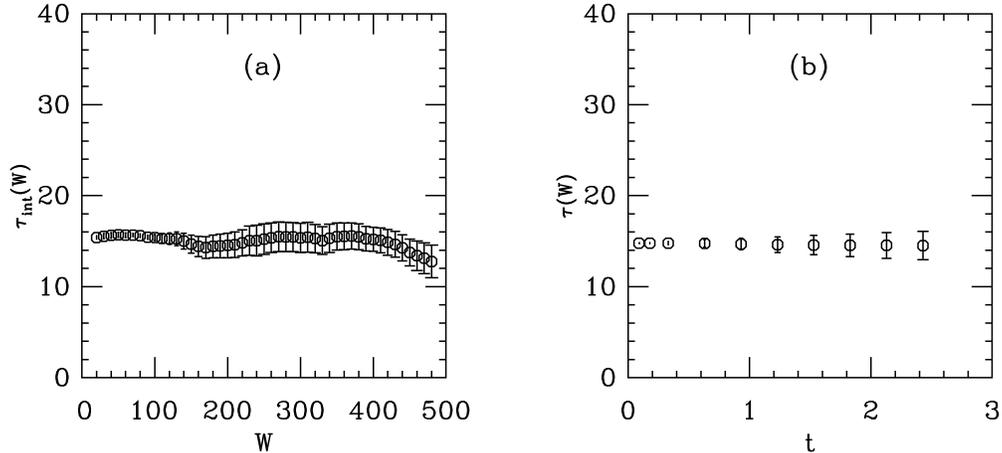}
\end{center}
\caption{Integrated auto-correlation  times for an $SU(4)$ data set, $\beta=10.2$, $\kappa=0.127$.
(a) $\tau_{int}(W)$ vs $W$ in molecular dynamics time units, at flow time $t=2.0$.
 (b) $\tau_{int}(W)$
for $W=20$ lattices (or $W=200$ molecular dynamics time units) for a set of flow values $t$.
\label{fig:int}}
\end{figure}

All of these data sets produce similar results. I show pictures from one data set, an $SU(4)$ 
gauge group with $\beta=10.2$, $\kappa=0.127$. Panel (a) of Fig.~\ref{fig:int} shows the integrated 
auto-correlation time
for $t^2E(t)$ as a function of $W$, measured in molecular dynamics time units (rescaled from data sets
spaced ten molecular dynamics units apart). Panel (b) shows $\tau_{int}(W=200)$ for a scan of flow time
values.
With a spacing of 10 molecular dynamics units between saved lattices, if an auto-correlation time were
less than
10 molecular dynamics units, it would be hard to observe.

Finally, there is the determination of $t_0$ (or of $C(N_c)$ itself). Here the issue is that on each
lattice, 
data at all values of flow times $t$ are correlated simply because later flow time data are constructed
by
processing earlier flow time data. I dealt with this by doing a jackknife analysis, basically along the
lines
of the ones done by Ref.~\cite{Bazavov:2015yea}. The analysis displayed in Fig.~\ref{fig:int} suggests
doing the jackknife eliminating sets of lattices whose length is longer than the integrated
auto-correlation time.
 This is two successive lattices for $\tau_{int}=20$ molecular dynamics time units.
I varied the size of the cut; even eliminating 10 successive lattices from the jackknife 
(100 molecular dynamics
time units) generally resulted in only a 20 per cent rise in the quoted uncertainty.

Two sets of numbers are needed, values of $C(N_c)$ at a fixed ratio of $t_0/r_1^2$,
and values of $t_0$ at an input $C(N_c)$. These values are determined by a fit to a small set of points
roughly centered around the fit value to a linear dependence ($t^2\svev{E(t)}=c_0 + c_1 t$)
 followed
by a linear interpolation to the desired value. These results were collected and the jackknife produced
the
numbers quoted in the table. I varied the range of the fit and the number of points kept; as long as the 
central values lie within the range of points kept, their values are insensitive to the fit range.

\section{Results}

\subsection{$C(N_c)$ vs $N_c$}

\begin{figure}
\begin{center}
\includegraphics[width=0.5\textwidth,clip]{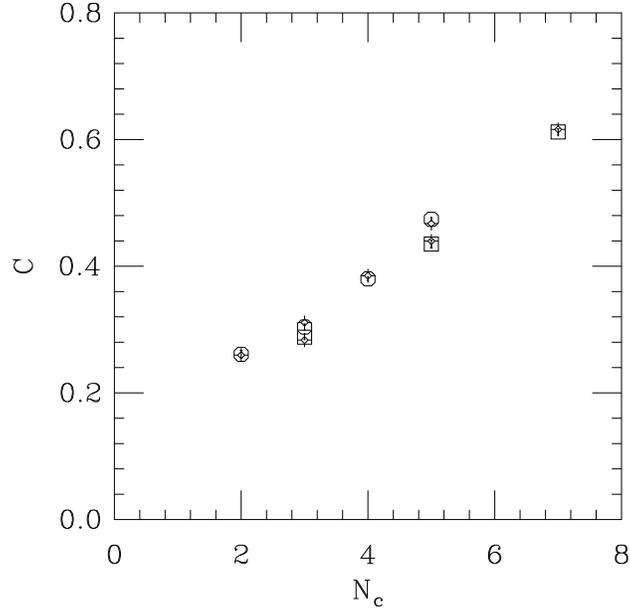}
\end{center}
\caption{Tuning factor $C(N_c)$ from \protect{\ref{eq:flow}}, matching
$\sqrt{t_0}/r_1=0.46$. Octagons are dynamical fermion data while squares 
are quenched. The fancy diamonds are a fit to both data sets described in the text.
\label{fig:cvsnc}}
\end{figure}

L\"uscher suggested taking $C(N_c)=0.3$ for $N_c=3$ QCD. The resulting $t_0$ has been evaluated
by many groups\cite{c1,c2,c3,c4,c5,Bazavov:2015yea}, $\sqrt{t_0}=0.14$ fm in $N_f=3$ QCD.
 (The quantity is actually known to four digits.)
Let us keep the ratio $\sqrt{t_0}/r_1$ 
fixed, $\sqrt{t_0}/r_1=0.46$,   while varying $N_c$, and ask how $C(N_c)$ is changed.
 Fig.~\ref{fig:cvsnc} shows data from quenched
$SU(N_c)$ simulations with $N_c=3$, 5, 7 \cite{DeGrand:2012hd}, and data
 from $N_f=2$ dynamical fermion simulations
with $N_c=2$, 3, 4, 5 \cite{DeGrand:2016pur}.
(Error bars in the figure are dominantly from the 
uncertainty in $r_1$.) The data are tabulated in Table \ref{tab:c0}.
 The dynamical fermion data are at roughly constant pseudoscalar to vector mass ratio, 
so they are matched in fermion mass.
The gauge couplings and fermion hopping parameters are $(\beta,\kappa)=(1.9,0.1295)$,
 (5.4, 0.127), (10.2, 0.1265) and (16.4,0.1265), for $N_c=2$, 3, 4, and 5,
 from the data sets of Ref.~\cite{DeGrand:2016pur}.  $C(N_c)$ clearly varies linearly with $N_c$.
It is not a pure linear dependence; $C(N_c) = A_1 N_c + A_0 + ...$ and the $A_0$ and higher
order contributions are due to $1/N_c$ corrections canceling
the leading $N_c$ dependence in Eq.~\ref{eq:scale}. 
 Presumably, the higher order corrections are also $N_f$ dependent.

I have not found a fit with a chi-squared per degree of freedom which is near unity. The
figure shows a one attempt: I fit all the data (quenched and $N_f=2$ to 
\bee
C(N_c,N_f)= c_1 N_c + c_2 N_f + \frac{c_3}{N_c}+c_4.
\ee
The fit has a $\chi^2$ of 11.6 for 3 degrees of freedom; 
$c_1=0.096(3)$,
$c_2=0.014(2)$,
$c_3=0.0267(46)$,
$c_4= -0.093(26)$.

Finally, the authors of Ref.~\cite{Ce:2016awn} use
\bee
 C(N_c)= 0.3 \left( \frac{3}{8} \frac{N_c^2-1}{N_c}\right)
\label{eq:ce}
\ee
to match scales in their quenched calculation of the topological susceptibility. This absorbs all the leading
factors of $N_c$ in the quenched versions of Eqs.~\ref{eq:pertflow}-\ref{eq:oneloop}
(or, said alternatively, makes an all-orders ansatz for its $N_c$ counting), 
while fixing the $N_c=3$ value to $C(3)=0.3$.
This seems to over-estimate the slope of $C(N_c)$ versus $N_c$, when compared to $r_1$, for the $N_f=2$ data sets.
It would give $C(N_c=7)=0.77$.

\begin{figure}
\begin{center}
\includegraphics[width=0.5\textwidth,clip]{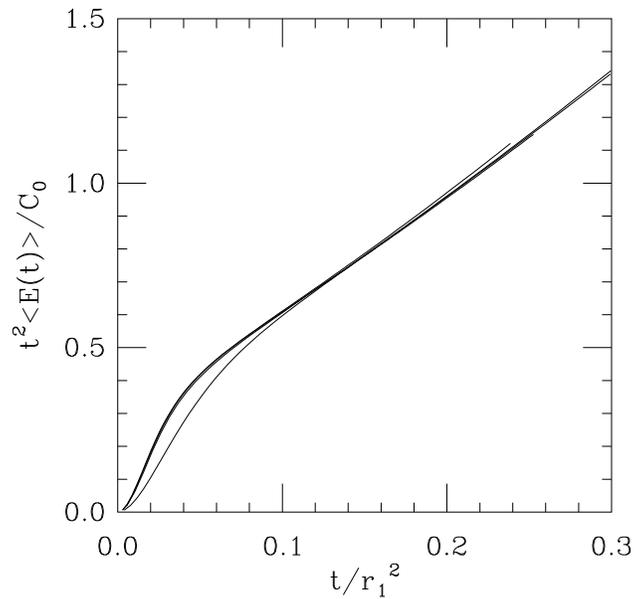}
\end{center}
\caption{Plots of $t^2\svev{E(t)}$ scaled by $N_c$-dependent constants, as a function of $t/r_1^2$.
The data sets and constants are:
(SU(2): $\beta=1.9$, $\kappa=0.1295$, $C_0=0.26$);
(SU(3): $\beta=5.4$, $\kappa=0.127$, $C_0=0.3$);
(SU(4): $\beta=10.2$, $\kappa=0.1265$, $C_0=0.38$);
(SU(5): $\beta=16.4$, $\kappa=0.1265$, $C_0=0.47$).
The $SU(2)$ curve is the slightly discrepant one at small $t$.
\label{fig:cvsncallr1}}
\end{figure}

I conclude this section by remarking that matching $C(N_c)$ by taking one value of $t/r_1^2$ to be an $N_c$ 
independent constant produces a match of lattice data at different $N_c$'s across a wide range of $t$.
This is displayed in Fig.~\ref{fig:cvsncallr1}.

\subsection{ $t_0$ vs $m_{PS}^2$}

\begin{figure}
\begin{center}
\includegraphics[width=0.6\textwidth,clip]{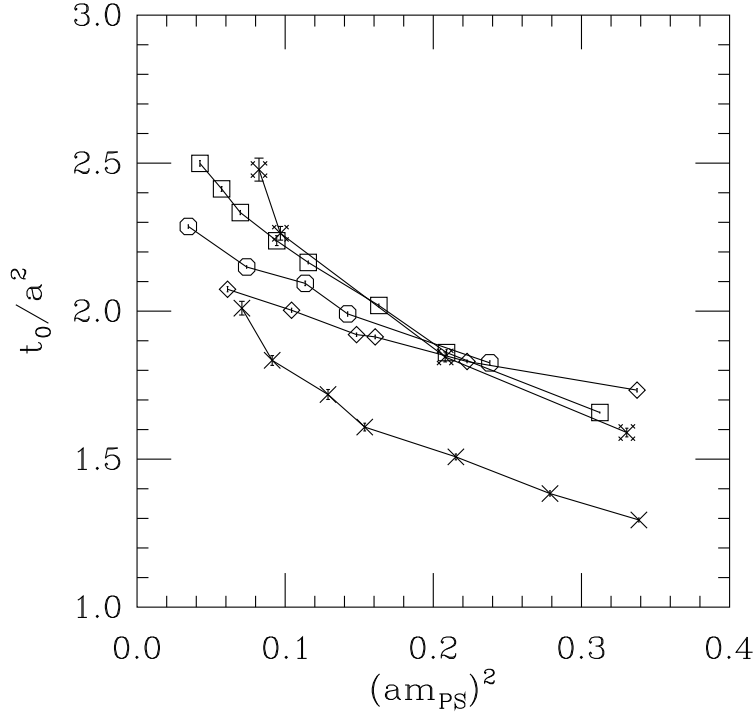}
\end{center}
\caption{The quantity $t_0/a^2$ versus squared pseudoscalar mass in lattice units, $(am_{PS})^2$,
for $N_c=2$ (crosses for $\beta=1.9$, fancy crosses for $\beta=1.95$),
 3 (squares), 4 (octagons), and 5 (diamonds).
\label{fig:t0vsmpi2}}
\end{figure}

I next fix the value of $C(N_c)$ and collect data at many values of the quark mass,
using the data sets of Ref.~\cite{DeGrand:2016pur}. 
I evaluate $t_0$ using the values of $t^2\svev{E(t)}$ which match length scales,
as shown in Fig.~\ref{fig:cvsnc}.
They are $C(N_c)=0.26$, 0.3, 0.38, and 0.47 for $N_c=2$, 3, 4, and 5.
 The data is tabulated in Table \ref{tab:t0}.
With this data, I ask, 
can we observe the fermion mass dependence of $t_0$  predicted by the
chiral Lagrangian analysis of B\"ar and Golterman  \cite{Bar:2013ora}?
They write an expansion for $E(t)$ in terms of the characteristic
length scale for a chiral Lagrangian,
\bee
E(t) = c_1 f_{PS}^4 + \dots + c_3 f_{PS}^2 \Tr (\chi^\dagger U + U^\dagger \chi) + \dots
\label{eq:first}
\ee
where $f_{PS}$ is the pseudoscalar decay constant, $U$ is the usual exponential of the
Goldstone boson field, $\chi$ is proportional to the fermion mass or to the squared
pseudoscalar mass $m_{PS}^2$,
and the $c_i$'s are a set of dimensionless coefficients.
They then predict
\bee
t_0(m_{PS}) = t_0(0)( 1 + k_1 \frac{m_{PS}^2}{f_{PS}^2} + k_2 \frac{m_{PS}^4}{f_{PS}^4}\log(\frac{m_{PS}^2}{\mu^2} )
+ k_3 ( \frac{m_{PS}^2}{f_{PS}^2} )^2 + \dots)
\label{eq:bg}
\ee
 where $t_0(0)$ is the value of the flow parameter at zero mass,
The $k_i$'s are also dimensionless constants, ratios of the $c_i$'s.
Judging from the quality of the data in Ref.~\cite{DeGrand:2016pur}, it should be possible to observe the
leading (proportional to $k_1$) mass dependence in this expression.
 The result is shown in Fig.~\ref{fig:t0vsmpi2}. There is a definite,
more or less linear, dependence of $t_0$ on the squared mass,
for all $N_c$'s. The slope  flattens as $N_c$ rises.

The flattening of the slope follows the naive expectation that fermions affect gauge observables less and less 
as $N_c$ rises. It also tells us a bit more. In Eqs.~\ref{eq:first} and \ref{eq:bg} the constants $c_i$
and $k_i$ are dimensionless, but of course this does not say anything about how the higher order terms
$c_3$ or $k_1$ scale with $N_c$.

Data from several $N_c$'s allows us to  say something about $k_1$.
 The pseudoscalar decay constant scales as $\sqrt{N_c}$. How does $k_1$ depend on $N_c$?  We can look at that
behavior by rescaling the data.
Eq.~\ref{eq:bg} can be rewritten as
\bee
\frac{t_0(m_{PS})}{t_0(0)} -1 = \frac{k_1}{f_{PS}^2}m_{PS}^2 + \dots.
\ee
I observe that $k_1/f_{PS}^2$ scales like $1/N_c$. To see if that expectation holds, plot
the scaled quantity $N_c(t_0(m_{PS})/t_0(0) -1)$ versus $m_{PS}^2$ and look for a common slope.

\begin{figure}
\begin{center}
\includegraphics[width=0.6\textwidth,clip]{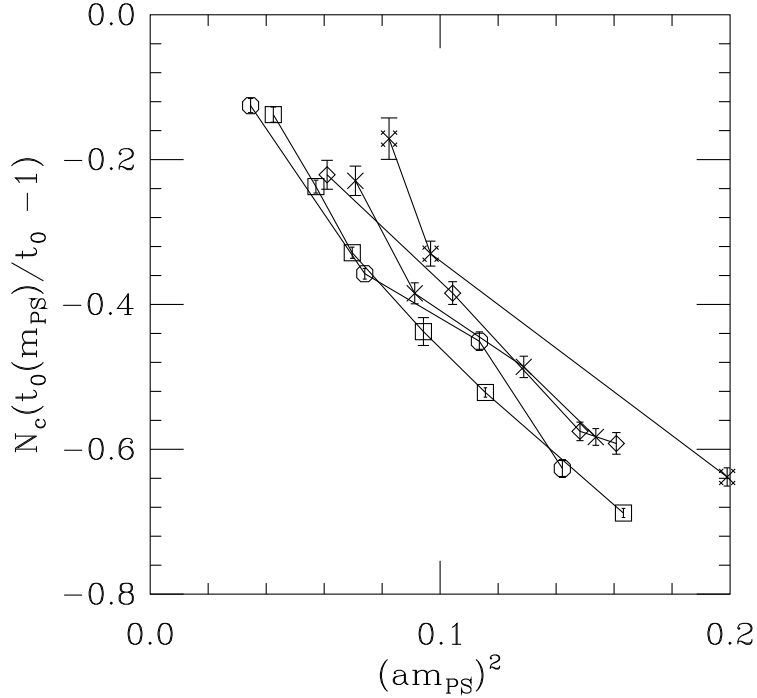}
\end{center}
\caption{The shifted quantity $N_c(t_0(m_{PS})/t_0 -1)$, 
 versus squared pseudoscalar mass in
lattice units, $(am_{PS})^2$,
for $N_c=2$ (crosses for $\beta=1.9$, fancy crosses for $\beta=1.95$),
 3 (squares), 4 (octagons), and 5 (diamonds).
\label{fig:nct0vsmpi2s}}
\end{figure}

 Fig.~\ref{fig:nct0vsmpi2s} shows this.
B\"ar and Golterman say that their formula is applicable for flow times much smaller than the square
of the pion wavelength. With $t_0 \sim 2-2.5a^2$ it seems appropriate to concentrate on $(am_{PS})^2<0.2$ or so,
and that is what is shown in the figure. The intercept $t_0$ is determined 
 by doing a quadratic fit of $t_0(m_{PS})$, $t_0(m_{PS}) = t_0(0) + A (am_{PS})^2 + B (am_{PS})^4$.
The plot uses $t_0(0)=2.27$, 2.71, 2.62, 2.36 and 2.17 for $SU(2)$ $\beta=1,9$, $SU(2)$ $\beta=1.95$, 
$SU(3)$, $SU(4)$, and $SU(5)$.
 Data for different $N_c\ge 3$ seems to behave similarly -- a linear dependence on $m_{PS}^2$
with an $N_c$ - independent slope. This says that $k_1$ is a constant, independent of $N_c$.
(Linear fits to the points shown in the figure give slopes $N_c k_1/f_{PS}^2= - 3.7(2)$ and -3.1(2) for the
$\beta=1.9$ and 1.95 $SU(2)$ points, -4.3(1) for $SU(3)$, -4.4(2) for $SU(4)$, and -3.9(2) for $SU(5)$.)

This result has a more mundane large $N_c$ origin. $E(t)$ is dominantly a gluonic observable,
$\svev{E(t)} \propto \svev{g^2 G^2)}$ (re-inserting a factor of $g^2$ as compared to Eq.~\ref{eq:def}).
$\svev{G^2}$ is also a gluonic observable, which scales as $N_c^2$. (Think of it as a closed gluon loop.) 
 The coupling scales as $g^2=\lambda/N_c$ for `t Hooft
coupling $\lambda$.   Thus,
 $\svev{E(t)}\propto N_c$ at fixed $\lambda$. This is the scaling
for $C(N_c)$ seen in Fig.~\ref{fig:cvsnc}.
 Because $f_{PS}$ scales as $\sqrt{N_c}$,
$c_1$  in Eq.~\ref{eq:first} must scale as $1/N_c$, and then $c_1 f_{PS}^4 \propto N_c$. The second term in Eq.~\ref{eq:first}
is a fermionic contribution to a gluonic observable, which is a $1/N_c$ effect: that is,
$(c_3 f_{PS}^2)/(c_1 f_{PS}^4) \propto 1/N_c$, or $k_1=c_3/c_1\propto N_c^0$. (Think of breaking the gluon
loop into a $q \bar q$ pair: this costs a factor of $g^2$ while leaving the double-line color counting
$N_c^2$ unchanged. Replacing $g^2$ by $\lambda/N_c$ gives a $1/N_c$ suppression.)
This is what Fig.~\ref{fig:nct0vsmpi2s} shows.

Note that the only parts of Eq.~\ref{eq:bg} which are unambiguously ``fermionic'' rather than
``gluonic,'' and which are accessible to simulation, are the terms with explicit quark mass (or
$m_{PS}$) dependence.

We would expect $N_c=2$ to be an outlier.
 The pattern of chiral
symmetry breaking is different for $SU(2)$  than for $N_c\ge 3$ since the fermions live in pseudo-real representations.
Generally, that means that the coefficients in a chiral expansion are different from the
usual factors appropriate to complex representations. Nevertheless, the plots of $t_0$ versus mass show 
empirically that the value of $k_1$ does not seem to be very different.

\section{Conclusions}
In this note I discussed the $N_c$ dependence of the flow scale $t_0$ and compared it to simple theoretical expectations.
I observed that in order to match the $t_0$ scale to that of other gluonic observables 
 it was necessary to scale $t^2\svev{E(t)}$ in a particular way with $N_c$.
(I used the Sommer parameter $r_1$,
derived from the heavy quark potential.)
I also observed the decoupling of $t_0$, a gluonic observable, from fermionic degrees of freedom, as $N_c$ grows.
Measurements of $t_0(m_{PS}^2)$ at several values of $N_c$ are the closest one can come to observing the 
$1/f_{PS}^2$ in the B\"ar-Golterman formula.

In QCD, the flow time $t_0$ is presently the quantity of choice for scale setting, and one would expect 
that it would find use in simulations of other confining and chirally broken systems. Researchers
who use it will discover that the dependence of $t^2 \svev{E(t)}$ on $t$ will be different for their system
than for $N_c=3$ QCD. An analysis similar to the one described here might allow them to justify 
some particular choice for
$C$. A useful part of the analysis of any new model is to ask ``how is it different from real world QCD?''
Part of the answer to this question involves the analysis of Monte Carlo data, and a scale choice is 
a necessary part of this analysis. A comparison of a new system with QCD might involve matching the scale
 choice used
for the new system with the one used for QCD, which would require an analysis similar to the one done here.

In addition, there is more to the analysis of a new system than Monte Carlo data. It is often useful
to have a model, which can hint at results which have not yet been computed on the lattice, or which may not
be accessible to the lattice. (Large $N_c$ counting is an example of such a model.) However, models
typically are incomplete.
Some observed behavior might have a simple and unexpected source (given by large $N_c$ counting, for example),
but it may not be something which can be completely justified from first principles.
It is always useful  to verify and confirm assumptions and common lore,
in a sound and reliable way.

\begin{acknowledgments}
I acknowledge important conversations with R.~Sommer   and B.~Svetitsky.
I thank the theory group at DESY-Zeuthen for its hospitality.
This work was supported  by the U.~S. Department of Energy, under grant DE-SC0010005.
Some computations were performed on the University of Colorado cluster.
Additional computations were done on facilities of the USQCD Collaboration at Fermilab,
which are funded by the Office of Science of the U.~S. Department of Energy.
The computer code is based on the publicly available package of the
 MILC collaboration~\cite{MILC}. The version I use was originally developed by Y.~Shamir and
 B.~Svetitsky.
\end{acknowledgments}

\begin{table}
\begin{tabular}{c c c c}
\hline
$\kappa$  &  $(a\,m_{PS})^2$ & $t_0/a^2$ & N \\
\hline
$SU(2)$ $\beta=1.9$ & $C=0.26$ & & \\
\hline
0.1280 & 0.339(2) & 1.295(7) & 40 \\
0.1285 & 0.279(3) & 1.384(9) & 40 \\
0.1290 & 0.215(3) & 1.508(10) & 40 \\
0.1295 & 0.154(3) & 1.608(13) & 40 \\
0.1297 & 0.129(2) & 1.718(17) & 40 \\
0.1300 & 0.091(3) & 1.833(16) & 40 \\
0.1302 & 0.071(3) & 2.010(23) & 40 \\
\hline
$SU(2)$ $\beta=1.95$ &  $C=0.26$& & \\
\hline
0.1270 & 0.331(3) & 1.590(15) & 40 \\
0.1280 & 0.208(2) & 1.845(17) & 40 \\
0.1290 & 0.097(2) & 2.263(23) & 40 \\
0.1292 & 0.082(2) & 2.478(39) & 40 \\
\hline
$SU(3)$ $\beta=5.4$ &  $C=0.3$&  & \\
\hline
0.1250 & 0.312(2) & 1.657(3) & 500 \\
0.1260 & 0.209(1) & 1.860(10) & 100 \\
0.1265 & 0.163(2) & 2.019(6) & 500 \\
0.1270 & 0.116(2) & 2.165(6) & 500 \\
0.1272 & 0.094(2) & 2.238(17) & 100 \\
0.1274 & 0.070(2) & 2.333(7) & 500 \\
0.1276 & 0.057(1) & 2.413(8) & 500 \\
0.1278 & 0.042(1) & 2.500(9) & 500 \\
\hline
$SU(4)$ $\beta=10.2$ &  $C=0.38$& & \\
\hline
0.1252 & 0.238(2) & 1.826(7) & 90 \\
0.1262 & 0.142(1) & 1.990(7) & 90 \\
0.1265 & 0.114(1) & 2.094(8) & 100 \\
0.1270 & 0.074(1) & 2.149(4) & 500 \\
0.1275 & 0.035(1) & 2.286(6) & 500 \\
\hline
$SU(5)$ $\beta=16.4$ &  $C=0.47$& & \\
\hline
0.1240 & 0.338(1) & 1.733(5) & 90 \\
0.1252 & 0.223(1) & 1.830(5) & 90 \\
0.1258 & 0.161(2) & 1.913(6) & 90 \\
0.1260 & 0.148(1) & 1.920(6) & 90 \\
0.1265 & 0.104(1) & 2.003(7) & 90 \\
0.1270 & 0.061(1) & 2.074(9) & 90 \\
\hline
 \end{tabular}
\caption{ $N_f=2$ dynamical fermion data plotted in the figures. The column labeled by $N$ gives the number
of lattice analyzed for $t_0$. The data is that of Ref.~\protect{\cite{DeGrand:2016pur}}.
Pseudoscalar masses are reproduced for convenience.
\label{tab:t0}}
\end{table}

\begin{table}
\begin{tabular}{c c c c}
\hline
$N_c$ & ${\beta}$ & $\kappa$ & $C(N_c)$ \\
\hline
3(Q)  & 6.0175 & & 0.288(4) \\
5(Q) & 17.5 &  & 0.435(6)  \\
7(Q) & 34.9 & & 0.612(5)  \\
\hline
2 & 1.9 & 0.1295 & 0.261(6) \\
3 & 5.4 & 0.127 & 0.305(5) \\
4 & 10.2 & 0.1265 & 0.380(3) \\
5 & 16.4 & 0.1265 & 0.474(3) \\
\hline
 \end{tabular}
\caption{ Data in Fig.~\protect{\ref{fig:cvsnc}}, $C(N_c)$ at $\sqrt{t}/r_1=0.46$.  ``Q'' labels quenched data.
\label{tab:c0}}
\end{table}


\end{document}